\documentclass[11pt]{article}

\usepackage{amsthm}
\usepackage{thmtools,thm-restate}
\usepackage{fullpage}
\usepackage{mathpazo}
\usepackage{xfrac}
\usepackage{setspace}

\usepackage{nth}
\usepackage{intcalc}
\usepackage{xstring}

\usepackage{ifpdf}
\ifpdf
\else
\usepackage[quadpoints=false]{hypdvips}
\fi

\newcommand{\ECCCurl}[2]{http://eccc.hpi-web.de/report/\ifnumcomp{#1}{>}{93}{19}{20}#1/#2/}

\newcommand{\shortECCC}[2]{\texttt{\href{http://eccc.hpi-web.de/report/\ifnumcomp{#1}{>}{93}{19}{20}#1/#2/}{eccc:TR#1-#2}}}

\newcommand{\parseECCC}[1]{% Takes a string of the form TRxx/xxx or
\StrSubstitute{#1}{TR}{}[\tmpstring]%
\IfSubStr{\tmpstring}{/}{ %assuming string is of the form TRxx/xxx
\StrBefore{\tmpstring}{/}[\ecccyear]%
\StrBehind{\tmpstring}{/}[\ecccreport]%
}{% assuming string is of the form TRxx-xxx
\StrBefore{\tmpstring}{-}[\ecccyear]%
\StrBehind{\tmpstring}{-}[\ecccreport]%
}%
\shortECCC{\ecccyear}{\ecccreport}}

\usepackage{commons}
\usepackage[usenames,dvipsnames]{color}
\RequirePackage[colorlinks=true]{hyperref}
\hypersetup{
  linkcolor=[rgb]{0.3,0.3,0.6},
  citecolor=[rgb]{0.2, 0.6, 0.2},
  urlcolor=[rgb]{0.6, 0.2, 0.2}
}

\onehalfspace

\title{New Lower Bounds against Homogeneous Non-Commutative Circuits}

\author{Prerona Chatterjee
  \footnote{Tel Aviv University, Israel. This work was done while the author was a postdoctoral researcher at the Institute of Mathematics of the Czech Academy of Sciences, Prague and was supported by the Czech Science Foundation GA{\v C}R grant 19-27871X. Email: prerona.ch@gmail.com}
\and 
Pavel \Hrubes
  \footnote{Institute of Mathematics of the Czech Academy of Sciences, Prague. This work was supported by Czech Science Foundation GA{\v C}R grant 19-27871X. Email: pahrubes@gmail.com.}
}

\declaretheorem[numberlike=equation]{theorem}
\declaretheorem[numberlike=equation]{lemma}
\declaretheorem[numberlike=equation]{claim}
\declaretheorem[numberlike=equation]{proposition}
\declaretheoremstyle[bodyfont=\it,qed=$\lozenge$]{defstyle} 
\declaretheorem[numberlike=equation,style=defstyle]{remark}
\declaretheorem[name=Open Problem]{openproblem}

\newcommand{\ckt}{\textbf{$\mathcal{C}$}}
\newcommand{\wt}{\operatorname{\mathsf{wt}}}
\newcommand{\spn}{\operatorname{\mathsf{span}}}
\renewcommand{\fam}{\textbf{$\mathcal{F}$}}

\newcommand{\dB}{\mathsf{B}}
\newcommand{\esym}[2]{\mathrm{OS}_{#1}^{#2}}

\renewcommand{\phi}{\varphi}
\renewcommand{\epsilon}{\varepsilon}

\begin{document}
\maketitle

\begin{abstract}  
  We give several new lower bounds on size of homogeneous non-commutative circuits. We present an explicit homogeneous bivariate polynomial of degree $d$ which requires homogeneous non-commutative circuit of size $\Omega(d/\log d)$. For an $n$-variate polynomial with $n>1$, the result can be improved to $\Omega(nd)$, if $d\leq n$,  or $\Omega(nd \frac{\log n}{\log d})$, if $d\geq n$.
%-- previously, the best lower bound known was $\Omega(n\log d)$   
  Under the same assumptions, we also give a quadratic lower bound for the ordered version of the central symmetric polynomial.   
\end{abstract}

\section{Introduction}

%\an{Please move macros to the file, delete redundant commands, and do not number by section}

Arithmetic Circuit Complexity aims to categorize  polynomials according to how hard they are to compute in  algebraic models of computation.
The most natural model is that of an arithmetic circuit: starting from variables or constants, the circuit computes new polynomials by means of addition and multiplication operations. 
The question is how many of these operations are needed. The most challenging problem is to prove super-polynomial lower bounds against arithmetic circuits computing a low-degree polynomial.
This is known as the $\VP$ vs $\VNP$ problem and is the algebraic analogue of the famed $\P$ vs. $\NP$ question.  
The classical result of Baur and Strassen \cite{Str73, BS83} gives an $\Omega(n\log d)$ lower bound for an $n$ variate polynomial of degree $d$. 
A variety of lower bounds has since been obtained by imposing various restrictions on the computational model - e.g., arithmetic formulas or monotone circuits \cite{Kal85, Val80}. 
But the result of Baur and Strassen remains the strongest lower bound on unrestricted arithmetic circuits. 

In this paper, we are interested in the non-commutative setting where multiplication does not multiplicatively commute. 
Starting with the seminal works of Hyafil \cite{Hya77} and Nisan \cite{Nis91}, non-commutative circuits are a well-studied object. 
The lack of commutativity is a severe limitation of the computational power which makes the task of proving circuit lower bounds apparently easier. 
Nisan gave an exponential lower bound for non-commutative formulas whereas, commutatively, the best bound is only quadratic \cite{Kal85, CKSV22}. 
Since then, it seemed that exponential non-commutative circuit lower bounds are just around the corner.  
Recently, Limaye, Srinivasan and Tavenas \cite{TLS22} proved such a lower bound in the \emph{homogeneous, constant depth} setting.
They showed that any constant depth $\Delta$ non-commutative homogeneous circuit for the \emph{iterated matrix multiplication polynomial} over $n$ variables of degree $d$ must have size $n^{\Omega(d^{\frac{1}{\Delta}})}$.
However for general circuits, even in the non-commutative setting, the strongest lower bound remains $\Omega(n\log d)$ \cite{Str73, BS83}. 

We improve this lower bound to $\Omega(nd/\log d)$ under the additional assumption that the non-commutative circuit is also homogeneous (see \autoref{sec:preliminaries} for definition). 
Non-commutatively, this is already interesting if $n=2$: we obtain a bivariate polynomial of degree $d$ which requires circuit size nearly linear in $d$. 
It is well-known that a (commutative or not) circuit computing a homogeneous polynomial of degree $d$ can be converted to an equivalent homogeneous circuit with at most a $d^2$ increase in size (see, e.g., \cite{HWY11}). 
Hence, homogeneity is not a serious restriction if either $d$ is small or if one is after a super-polynomial lower bound -- as in the $\VP$ vs $\VNP$ problem. 
However, our results fall in neither category and we do not know how to remove the homogeneity restriction.
Nevertheless, we strongly believe that it can be removed and  non-commutative circuit lower bounds are just around the corner. 

\section{Notation and preliminaries}\label{sec:preliminaries}

Let $\F$ be a field. A \emph{non-commutative polynomial} over $\F$ is a formal sum of products of variables and field elements. We assume that the variables do not multiplicatively commute, whereas they commute additively, and with elements of $\F$. 
The ring of non-commutative polynomials in variables $x_1,\dots, x_n$ is denoted $\F\inangle{x_1,\dots,x_n}$. 
A polynomial is said to be \emph{homogeneous} if all monomials with a non-zero coefficient in $f$ have the same degree. 

A \emph{non-commutative arithmetic circuit} $\ckt$ is a directed acyclic graph as follows.
Nodes (or gates) of in-degree zero are labelled by either a variable 
or a field element in $\F$.
All the other nodes have in-degree two and they are labelled by 
either $+$ or $\times$.
The two edges going into a gate labelled by $\times$ are labelled by \emph{left} and \emph{right} to indicate the order of multiplication. Gates of in-degree zero will be called \emph{input} gates; gates of out-degree zero will be called \emph{output} gates.  

Every node in $\ckt$ computes a non-commutative polynomial in the obvious way. We say that $\ckt$ computes a polynomial $f$ if there is a gate in $\ckt$ computing $f$ (not necessarily an output gate). 
$\ckt$ will be called \emph{homogeneous} if every gate in $\ckt$ computes a homogeneous polynomial. 
Given a circuit $\ckt$, let $\widehat{\ckt}:= \setdef{f}{f \text{ is computed by some gate in } \ckt}$.

A product gate will be called \emph{non-scalar}, if both of its inputs compute a non-constant polynomial. 
We define the \emph{size} of $\ckt$ to be the number of non-input gates in it, and the \emph{non-scalar size} of $\ckt$ to be the number of non-scalar product gates in it. 

Given integers $n_1,n_2$, $[n_1,n_2]$ is the interval $\set{n_1,n_1+1,\dots, n_2}$ and $[n]:=[1,n]$.  
%For any polynomial $f$, we denote its degree by $\deg(f)$ and use $\hom_i(f)$ to denote the $i$-th homogeneous component of $f$ for any $0 \leq i \leq \deg(f)$.

\bigskip
\noindent {\bf Note:} Unless stated otherwise, circuits and polynomials are assumed to be non-commutative and the underlying field $\F$ is fixed but arbitrary.

\section{Main results}

For univariate polynomials there is no difference between commutative and non-commutative computations.
Already with two variables, non-commutative polynomials display much richer structure.
There are $2^d$ monomials in variables $x_0,x_1$ of degree $d$ (as opposed to $d+1$ in the commutative world); so a generic bivariate polynomial requires a circuit of size exponential in $d$.  

Our first result is a lower bound that is almost linear in $d$. 
The hard polynomial is a bivariate monomial (a specific product of variables $x_0,x_1$).
 
\begin{restatable}{theorem}{linLBmonomial}\label{thm:lin_lb_monomial}
  For every $d>1$, there exists an explicit bivariate monomial of degree $d$ such that any homogeneous non-commutative circuit computing it has non-scalar size $\Omega(d/\log d)$.
\end{restatable}

\noindent In \autoref{rem:tight}, we point out a complementary $O(d/\log d)$ upper bound for every bivariate monomial. 
Note that commutatively every such monomial can be computed in size $O(\log d)$.

For $n$-variate polynomials, we obtain a stronger result (the hard polynomial is no longer a monomial).   

\begin{theorem}\label{thm:mainnd} For every $n,d>1$ there exists an explicit $n$-variate homogeneous polynomial of degree $d$ which requires a homogenous  non-commutative circuit of non-scalar size $\Omega(nd)$, if $d\leq n$, or $\Omega(nd \frac{\log n}{\log d})$, if $d\geq n$.  
\end{theorem}

\noindent \autoref{thm:lin_lb_monomial} and \autoref{thm:mainnd} are proved in \autoref{sec:monomial} and \autoref{sec:nd} respectively.

Given $0\leq d, n$, the \emph{ordered symmetric polynomial}, $\esym{n}{d}$, is the polynomial\footnote{Hence $\esym{n}{0}=1$ and $\esym{n}{d}=0$ whenever $d>n$.}
\[ 
 \esym{n}{d}(x_1,\dots,x_n) = \sum_{1\leq i_1 < \cdots < i_d \leq n} \inparen{\prod_{j=1}^{d} x_{i_j}}\,.
\]
It can be thought of as an ordered version of the commutative elementary symmetric polynomial. 
In \autoref{sec:esym}, we shall prove a lower bound for this polynomial.

\begin{restatable}{theorem}{quadLBcsym}\label{thm:quad_lb_csym}
  If $2\leq d\leq n/2$, any homogeneous non-commutative circuit computing $\esym{n}{d}(x_1,\dots,x_n)$ must have non-scalar size $\Omega(dn)$.
\end{restatable}

For the central ordered symmetric polynomial $\esym{n}{\floor{n/2}}$, the lower bound becomes $\Omega(n^2)$. 
We also observe that the known commutative upper bounds on elementary symmetric polynomials work non-commutatively as well.

\begin{proposition}\label{prop:ub}
  $\esym{n}{1}, \dots, \esym{n}{n}$ can be simultaneously computed by a non-commutative circuit of size $O(n\log^2n\log\log n)$, and by a homogeneous non-commutative circuit of size $O(n^2)$. 
\end{proposition}
 
The polylog factor in the proposition depends on the underlying field and can be improved for some $\F$s. 
Moreover, when measuring non-scalar size, one can obtain an $O(n\log n)$ upper bound if $\F$ is infinite -- this is tight by \cite{BS83}. 
 
The ordered symmetric polynomial can be contrasted with the truly symmetric polynomial 
\[ 
  S^k_n= \sum_{i_1, \dots, i_k\in [n]\hbox { distinct}} x_{i_1}\cdots x_{i_k}\,,
\]
Non-commutatively, already $S_{n}^n$ is as hard as the permanent \cite{HWY11} and is expected to require exponential circuits. 

\begin{remark} 
  A polynomial of degree $d$ can be uniquely written as $f=\sum_{k=0}^df^{(k)}$ where $f^{(k)}$ is homogeneous of degree $k$. 
  It is well-known that if $f$ has a circuit of size $s$, the homogeneous parts $f^{(0)},\dots, f^{(d)}$ can be simultaneously computed by a homogeneous circuit of size $O(sd^2)$ (this holds non-commutatively as well \cite{HWY11}). 
  Note that $\esym{n}{0},\dots, \esym{n}{n}$ are the homogeneous parts of $\prod_{i=1}^n (1+x_i)$ which has a circuit of a linear size. 
  \autoref{thm:quad_lb_csym} shows that in this case, homogenization provably costs a factor of the degree.      
\end{remark}

\section{Lower bounds against homogeneous non-commutative circuits}

Let us define the measure we use to prove our lower bounds.
Suppose $f \in \F \inangle{x_1, \ldots, x_n}$ is a homogeneous polynomial of degree $d$. 
Given an interval $J= [a,b]\subseteq [d]$, the polynomial $f^J$ is obtained be setting variables in position \emph{outside} of $J$ to one. 
More precisely, if $\alpha=\prod_{i=1}^d x_{j_i}$ is a monomial then $\alpha^J:= \prod_{i=a}^b x_{j_i}$, and the map is extended linearly so that $f^J= \sum_k c_k \alpha_k^J $ whenever $f= \sum_k c_k \alpha_k$. 
Given a non-negative integer $\ell$, let 
\[
  \fam_\ell(f) = \setdef{f^J}{J \subseteq [d] \text{ is an interval of length } \ell}.
\]
Given homogeneous polynomials $f_1,\dots, f_m$, our hardness measure is defined as 
\[
 \mu_\ell(f_1,\dots,f_m) := \dim(\spn(\bigcup_{i=1}^m \fam_\ell(f_i)))\,.
\]
Here, $\spn(\fam)$ denotes the vector space of $\F$-linear combinations of polynomials in $\fam$ and $\dim$ is its dimension. 

The following lemma bounds the measure in terms of circuit size. 

% \[\fam_\ell(\ckt) = \bigcup_{\substack{g \text{ : } g \text{ is computed}\\\text{at some gate in \ckt}}} \fam_\ell(g) \]

%\an{Note that it was *false* for $\ell=1$ when counting wires (and it remains so when we count non-input gates): $x_1, \dots, x_n$ have circuit of size 0 but $\mu_1=n$. }

\begin{lemma}\label{lem:meausure_ub_hom_nc_ckt}\label{lem:mu}
  Let $\ckt$ be a homogeneous circuit with $s$ non-scalar multiplication gates. Then for every $\ell\geq 2$, $\mu_\ell(\widehat {\ckt})\leq (\ell-1)s$. 
\end{lemma}

\begin{proof} 
  This is by induction on the size of $\ckt$. 
  If $\ckt$ consists of input gates only then $\fam_\ell(\widehat {\ckt})=\emptyset$, as we assumed $\ell\geq 2$ and $\widehat {\ckt}$ consists of linear polynomials. 

  Otherwise, assume that $u$ is some output gate of $\ckt$ and let $\ckt'$ be the circuit obtained by removing that gate. 
  If $u$ is a sum gate or a scalar product gate then 
  \[
    \mu_\ell(\widehat {\ckt})\leq  \mu_\ell(\widehat {\ckt'})\,.
  \]
  For if $u$ computes $f$  then $f=a_1f_1+a_2f_2$ for some constants $a_1, a_2$ and $f_1,f_2\in \widehat {\ckt'}$. 
  If $f$ has  degree $d$ then for every interval  $J\subseteq [d]$ of length $\ell$, $f^J=(a_1f_1+a_2f_2)^J= a_1f_1^J+a_2f_2^J\in \spn(\fam_\ell(\widehat{\ckt'}))$. 

  If $u$ is a non-scalar product gate computing $f=f_1\cdot f_2$ then 
  \[
    \mu_\ell(\widehat {\ckt})\leq \mu_\ell(\widehat {\ckt'})+ (\ell-1)\,.
  \]
  To see this assume $f_1,f_2$ have degrees $d_1$ and $d_2$ respectively, and let $J\subseteq [d_1+d_2]$ be an interval of length $\ell$. 
  If $J$ is contained in $[d_1]$, $f^J= (f_1f_2)^J= f_1^J f_2^\emptyset$ is a scalar multiple of $f_1^J$ and hence $f^J$ is contained in $\spn(\fam_\ell(\widehat {\ckt'}))$; similarly if $J$ is contained in $[d_1+1,d_2]$. 
  Otherwise, both $d_1$ and $d_1+1$ are contained in $J$. 
  But there are only $\ell-1$ such intervals.  
  Hence $\fam_\ell(\widehat {\ckt})$ contains at most $\ell-1$ polynomials outside of $\spn(\fam_\ell(\widehat {\ckt'}))$. 
 
  This means that $\mu_\ell$ increases only at product gates, and that it increases only by $\ell-1$ at such gates.
  Hence $\mu_\ell(\widehat {\ckt})\leq (\ell-1)s$.
\end{proof}

\begin{remark} 
  If $f$ has $n$ variables and degree $d$, the measure $\mu_\ell(f)$ can be at most the minimum of $d- (\ell-1)$ and $n^\ell$. 
  Hence, \autoref{lem:meausure_ub_hom_nc_ckt} can by itself give a lower of at most the order of $d\log n/\log d$. %Such a lower bound can be achieved for any $n\leq d$ and an explicit $f$ as in \autoref{sec:monomial}.  
\end{remark}

%On the other hand, we now show that the general statement of \autoref{lem:meausure_ub_hom_nc_ckt} can be used to prove an $\Omega(d/\log d)$ lower bound for a \emph{bivariate} monomial of degree $d$.

\subsection{Lower bounds for a single monomial}\label{sec:monomial}
Interestingly, \autoref{lem:mu} gives non-trivial lower bounds for $f$ being merely a product of variables. 
The simplest example is an $n$-variate product of a quadratic degree.

\begin{proposition} 
  Every homogeneous circuit computing $f=\prod_{i=1}^n\prod_{j=1}^n (x_ix_j)$ contains at least $n^2$ non-scalar product gates. 
\end{proposition}

\begin{proof} 
  This is an application of \autoref{lem:mu} with $\ell=2$. 
  The family $\fam_2(f)$ consists of all monomials $x_ix_j$. 
  Hence, $\mu_2(f)=n^2$. If $\ckt$ computes $f$, we have $\mu_2(\widehat {\ckt})\geq \mu_2(f)$ and hence $\ckt$ contains at least $n^2$ product gates.
\end{proof}

Another case of interest is a monomial in two variables, $x_0,x_1$, of degree $d$. 
Suppose $f= \prod_{i=1}^d x_{\sigma_i} $ where $\sigma= ( \sigma_1,\dots, \sigma_d)\in \{0,1\}^d $. 
Then $\mu_\ell(f)$ equals the number of distinct substrings of $\sigma$ of length $\ell$. 
Hence we want to find a $\sigma$ which contains as many substrings as possible. 
One construction of such an object is provided by the \emph{de Bruijn sequence} \cite{dB46}.

\paragraph{de Bruijn sequences}
  For a given $k$, a de Bruijn sequence of order $k$ over alphabet $A$ is a cyclic sequence $\sigma$ in which every $k$-length string from $A^k$ occurs exactly once as a substring.  
  Note that $\sigma$ must have length $|A|^k$.
  Furthermore, precisely  $k-1$ of the substrings  overlap the beginning and the end of the sequence and $\sigma$ contains $|A|^k -(k-1)$ substrings when viewed as an ordinary sequence. 
  de Bruijn sequences  are widely studied and, in particular, they exist.
  Moreover, efficient algorithms are known for constructing de Bruijn sequences (see, for example, \cite{Sww16} and its references).  
  In the case of binary alphabet $A=\{0,1\}$, this is especially so.
  We can start with a string of $k$ zeros. At each stage, extend the sequence by $1$, unless this results in a $k$-string already encounters, otherwise extend by $0$.    

  Given $d\geq 2$, let $\sigma$ be a binary de Bruijn sequence of order $\lceil \log_2 d\rceil $. 
  It has length $2^{\lceil \log_2 d\rceil} \geq d$. Define the polynomial 
  \[ 
    \dB_d(x_0,x_1):= \prod_{i=1}^d x_{\sigma_i}\,.
  \]
  The following implies the result of \autoref{thm:lin_lb_monomial}.

\begin{proposition}\label{prop:dB} 
  Every homogeneous circuit computing $\dB_d$ contains $\Omega(d/\log d)$ non-scalar product gates. 
\end{proposition}

\begin{proof}  
  This is an application of \autoref{lem:mu} with $\ell=\lceil \log_2 d\rceil$. 
  $[d]$ contains $d-{\ell-1}$ intervals of length $\ell$, all of which give rise to different substrings of $\sigma$. 
  The family $\fam_\ell(\dB_d)$ consists of $d-(\ell-1)$ different monomials and hence $\mu_\ell(\dB_d)=d-(\ell-1)$. 
  By the lemma, assuming $\ell>1$, a homogenous circuit for $\dB_d$ must contain $(d-(\ell-1))/ (\ell-1)=\Omega(d/\log d)$ product gates.  
\end{proof}

\begin{remark} \label{rem:tight}
  Using de Bruijn sequences over alphabet of size $n$, one can give an explicit monomial in $n>1$ variables and degree $d\geq n$ which requires homogeneous circuit of non-scalar size $\Omega(d\log n/ \log d)$.  
  This can also be deduced from \autoref{prop:dB} by viewing degree $k$ bivariate monomials as a single variable. 

  Conversely, every such monomial $\alpha$ can be computed in size $O(d\log n/ \log d)$ using multiplication gates only (such a computation is automatically homogeneous). 
  Indeed, we can first compute all monomials of degree at most $k$ by a circuit of size $O(n^{k+1})$ and then compute $\alpha$ using $\lceil d/k\rceil$ additional multiplication gates. 
  Choosing $k$ around $0.5\log_2d\log^{-1}_2n$ is sufficient. This also means the bound in \autoref{thm:mainnd} is tight.
\end{remark}

\subsection{Computing partial derivatives simultaneously}\label{sec:nd}

In order to obtain stronger lower bounds, we will translate the 
classical theorem of Baur and Strassen \cite{BS83} on computing partial derivatives to the non-commutative setting. 

We define partial derivative with respect to first position only, as follows. Given a polynomial $f$ and a variable $x$, $f$ can be uniquely written as $f=xf_0+f_1$ where no monomial in $f_1$ contains $x$ in the first position. 
We set $\partial_xf := f_0$. 

The proof of the following lemma is almost the same as the one of Baur and Strassen. An additional twist is added since we want the derivatives to be computed by a homogeneous circuit. 
This requires the generalization of homogeneity to allow arbitrary variable weights.
We emphasize that taking derivatives with respect to the first position is essential in the non-commutative setting.

\begin{lemma}\label{lem:computing-partials-simultaneously}
  Assume that $f \in \F\inangle{x_1, \ldots, x_n}$ can be computed by a homogeneous circuit of size $s$ and non-scalar size $s_{\times}$. 
  Then $\partial_{ x_1}f, \ldots, \partial_{x_n}f$ can be simultaneously computed by a homogeneous circuit of size $O(s)$ and non-scalar size $O(s_\times)$.
\end{lemma}

\begin{proof}
  Given $\vecw = (w_1, \ldots, w_n) \in \N^n$, let $w_i$ be the \emph{weight} of $x_i$ and let the weight of a monomial $\alpha= \prod_{j=1}^{d} x_{i_j}$ be defined as $\wt(\alpha) = \sum_{j=1}^{d} w_{i_j}$.
  A polynomial $f \in \F\inangle{x_1, \ldots, x_n}$ is said to be $\vecw$-homogeneous if every monomial in it has the same weight.  We call this the weight of $f$, denoted by $\wt(f)$.
  Furthermore we say that a circuit $\ckt$ is $\vecw$-homogeneous if every gate in it computes a $\vecw$-homogeneous polynomial.
  The weight of any node, $v$, in a $\vecw$-homogeneous circuit is defined to be the weight of the polynomial being computed by it.

  Note that if $(w_1, \ldots, w_n) = (1, \ldots, 1)$, then  $\vecw$-homogeneity coincides with the usual notion of homogeneity. %Furthermore, it is more convenient to measure circuit size by the number of edges in it (a..k.a. \emph{wires}) which differs from number of gates by at most a constant factor.   
  Therefore \autoref{lem:computing-partials-simultaneously} follows from the following claim.
  
  \begin{claim}
    For any $\vecw = (w_1, \ldots, w_n) \in \N^n$, if there is a $\vecw$-homogenous circuit that computes $f \in \F\inangle{x_1, \ldots, x_n}$ of size $s$ and non-scalar size $s_\times$, then there is a $\vecw$-homogeneous  circuit that computes $\mathbb{D}(f) = \set{\partial_{ x_1}f, \ldots, \partial_{ x_n}f}$ of size at most $5s$ and non-scalar size at most $2s_\times$.
  \end{claim}

  We prove this claim by induction on $s$. Recall that circuit size is measured by the number of non-input gates. 
  For the base case, $s=0$, the circuit only consists of leaves. The derivatives are then either $0$ or $1$ and can again be computed in zero size.  
  
  Assume $s > 0$.
  Let $\vecw = (w_1, \ldots, w_n) \in \N^n$ be arbitrarily fixed.
  Furthermore, suppose there is a $\vecw$-homogenous  circuit $\ckt$ that computes $f \in \F\inangle{x_1, \ldots, x_n}$ of size $s$.
  Choose a vertex $v$ in $\ckt$ such that both its children are leaves, and let $\widehat {v}$ be the polynomial it computes.  
  $\widehat {v}$ is a homogeneous polynomial in at most two variables and degree at most two;  w.l.o.g., we can also assume that $\widehat {v}$ is at least linear (otherwise $v$ could be replaced by a leaf).  
  
  Let $\ckt'$ be the circuit obtained from $\ckt$ by removing the incoming edges to $v$ and labelling the vertex $v$ with a new variable, say $x_0$. 
  Let us assign it weight $w_0:=\wt(\widehat {v})$.
  
  Let $f'$ be the polynomial computed by $\ckt'$.
  Then, $\mathbb{D}(f) = \set{\partial_{ x_1}f, \ldots, \partial_{ x_n}f}$ can be recovered from $\mathbb{D}(f') = \set{\partial_{ x_0}f', \partial_{ x_1}f', \ldots, \partial_{ x_n}f'}$ using the following version of chain rule:
  \[ 
    \partial_{x_k}f = (\partial_{x_k}f'+ \partial_{x_k}\widehat {v}\cdot  \partial_{x_0}f')|_{x_0: = \widehat {v}}\,. 
  \]
  Note that $\partial_{x_k}\widehat{v}$ is a variable or a constant, and that it is zero except for at most two of the $x_k$'s.

  Let us set $\vecw' = (w'_0, w_1, \ldots, w_n)$.
  Note that the weight of every vertex in $\ckt'$ is the same as the corresponding vertex in $\ckt$.
  Therefore, since $\ckt$ is $\vecw$-homogeneous, $\ckt'$ is $\vecw'$-homogeneous.
  Furthermore, $\ckt'$ has $s-1$ non-input gates and, by the inductive assumption, there is a $\vecw'$-homogeneous circuit $\mathcal{D}'$ of size $5(s-1)$ which computes $\mathbb{D}(f')$.
  Using $\mathcal{D}'$ and the chain rule above, we can construct a circuit with $5$ additional gates which computes $\mathbb{D}(f)$.
  The size of this circuit is at most $5(s-1) + 5 = 5s$ and is easily seen to be $\vecw$-homogeneous. 
  
  When counting non-scalar complexity, note that in the construction, only non-scalar product gates introduce non-scalar gates, and we always introduce at most two such gates.    
\end{proof}

We can now prove \autoref{thm:mainnd}. 

\begin{proof}[Proof of \autoref{thm:mainnd}] 
  Let $n,d$ be given with\footnote{If $d= 2$, $\esym{n}{2}$ satisfies the theorem; see \autoref{rem:standard}.} $n>1$, $d>2$. 
  Let $k$ be the smallest integer such that $n^k\geq n(d-1)$. Take a de Bruijn sequence $\sigma$ of order $k$ in alphabet $[n]$. 
  Take sequences $\sigma^1,\dots,\sigma^n\in [n]^{d-1}$ so that their concatenation $\sigma^1\dots \sigma^n$ is the initial segment of $\sigma$. 
  Define the polynomial 
  \[ 
    f= x_1\alpha_1+\dots+x_n\alpha_n\, \hbox{, where  }\alpha_i= \prod_{j=1}^{d-1}x_{\sigma^i_j} \,.
  \]

  Assume $f$ has a homogeneous circuit of non-scalar size $s$. 
  Then, by \autoref{lem:computing-partials-simultaneously}, $\alpha_1,\dots,\alpha_n$ can be simultaneously computed by a homogeneous circuit of size $s'=O(s)$. 
  We now apply \autoref{lem:mu} with $\ell=k$. 
  By construction, $\mu_k(\alpha_1,\dots, \alpha_n)= n(d-1- (k-1))=n(d-k)$. 
  This is because $\alpha_i^J$ are distinct monomials for different $i$'s and intervals of length $k$. 
  The lemma then gives $s'\geq n(d-k)/(k-1)$. 
  If $d\leq n$, we have $k=2$ and so $s'\geq n(d-2)$.  
  If $d>n$, we have $k\leq c_1 \log_2d/\log_2n$ and $d-k\geq c_2 d$, for some constants $c_1,c_2>0$.  
  Hence indeed $s'\geq \Omega(nd\frac{\log n }{\log d})$. 
\end{proof}

\subsection{Lower bound for ordered symmetric polynomials}

We now prove \autoref{thm:quad_lb_csym}. We first note: 

\begin{remark}\label{rem:standard} 
  $\esym{n}{2}$ requires $\Omega(n)$ non-scalar product gates (even in the commutative setting). 
  This can be proved by a standard partial derivatives argument as in \cite{NW97}. 
\end{remark}

Hence we can focus on degree $d>2$, in which case we give the following strengthening of \autoref{thm:quad_lb_csym}:

\begin{theorem}\label{thm:esym}
  If $1<k < n $, any homogeneous circuit computing $\esym{n}{k+1}(x_1,\dots,x_n)$
  requires  non-scalar size $\Omega(k(n-k))$.
\end{theorem}

\begin{proof} 
  Assume that a homogeneous circuit computes $f = \esym{n}{k+1}(x_1,\dots,x_n)$ using $s$ non-scalar product gates.  
  Then by \autoref{lem:computing-partials-simultaneously} there is a homogeneous  circuit of non-scalar size $O(s)$ which simultaneously computes $\set{\partial_{x_1}f, \ldots, \partial_{x_{n}}f}$.
  Let this circuit be $\ckt$.
  Then, by \autoref{lem:meausure_ub_hom_nc_ckt}, $\mu_2(\widehat {\ckt}) \leq O(s)$. 
  Note that
  \[
    \partial_{x_i}f = \esym{n-i}{k}(x_{i+1}, \ldots, x_n)\,.
  \]
  Let $f_{i,j}:=(\partial_{x_i}f)^{[j,j+1]}$.  
  We claim that the polynomials in  $F:=\setdef{f_{i,j}}{i \in [n- k], j \in [k-1]}$ are linearly independent. 
  This implies  that $\mu_2(\widehat{\ckt}) \geq (n-k)(k-1)$ and gives a lower bound of $\Omega(k(n-k))$ as required.
  
  We now prove that $F$ is indeed linearly independent. Consider the lexicographic ordering on $S:=[n-k]\times [k-1]$ defined by: 
  \[( i_0, j_0) < (i,j) \hbox{ iff } (j_0 > 
  j) \hbox{ or } (j_0=j  \hbox{ and } i_0< i)\,.\] 
  Let $(i_0,j_0)\in S$ be given. 
  Denote $\delta_{i_0,j_0}(g)$ the coefficient of 
  the monomial $x_{i_0+j_0}x_{n+j_0-k+1}$ in $g$. Then for every $(i,j)\in S$, 
  \begin{equation}\label{eq:ij}
    \delta_{i_0,j_0}(f_{i,j}) =
    \begin{cases}
      1 \text{\quad if $(i_0,j_0) = (i,j)$}\,\\
      0 \text{\quad if $(i_0,j_0) < (i,j)$}\,.
    \end{cases}
  \end{equation}

  To see (\ref{eq:ij}), assume that $\partial_{x_i}f$ contains $x_{n+j_0-k+1}$ in position $j+1$ in some monomial $\alpha$ with a non-zero coefficient.  
  The degree of $\alpha$ is $k$, and the positions $j+1,\dots, k$  need to be filled with variables from $x_{n+j_0-k+1},\dots,x_n$ in an ascending order. 
  There are $k-j$ such positions and $k-j_0$ such variables. 
  Therefore $j\geq j_0$. 
  Furthermore, if $j=j_0$, the last $k-j_0$ positions in $\alpha$ are uniquely determined as the variables $x_{n+j_0-k+1},\dots,x_n$ in that order. 
  Similarly, if $\partial_{x_{i}}f$ contains $x_{i_0+j_0}$ in position $j_0$ in some $\alpha$, the first $j_0$ positions must be filled with variables from $x_{i+1},\dots, x_{i_0+j_0} $. 
  Hence $i\leq i_0$, and in case of equality, the first $j_0$ positions are uniquely determined. 
  This means that $ \delta_{i_0,j_0}(f_{i,j})=0$ whenever $(i_0,j_0) < (i,j)$. 
  Furthermore, $\alpha:=\prod_{p=i_0+1}^{i_0+ j_0}x_{p}\prod_{p= n+j_0-k+1}^n x_p$  is the unique monomial in $f_{i_0,j_0}$ with $\delta_{i_0,j_0}(\alpha)=1$, concluding (\ref{eq:ij}).   
   
  Finally, assume for the sake of contradiction that there exists a non-trivial linear combination 
  \[
    \sum_{(i,j)\in S}  \gamma_{i,j} f_{i,j}=0\,.
  \]
  Let $( i_0,j_0)$ be the first pair in the lexicographic ordering with $\gamma_{i_0,j_0} \neq 0$. 
  Then we have
  \[
    0=\sum_{(i,j)\in S}  \gamma_{i,j} \delta_{i_0,j_0}( f_{i,j})=\gamma_{i_0,j_0} \delta_{i_0,j_0}(f_{i_0,j_0})+\sum_{(i,j)> (i_0,j_0)}  \gamma_{i,j} \delta_{i_0,j_0}(f_{i,j})\,.  
  \]
  Using (\ref{eq:ij}), the last sum is zero and  $\gamma_{i_0,j_0} \delta_{i_0,j_0}(f_{i_0,j_0})= \gamma_{i_0,j_0}=0$,  contrary to the assumption  $\gamma_{i_0,j_0}\not =0$.
\end{proof}

\section{Upper bounds for ordered symmetric polynomials}\label{sec:esym}

In \autoref{prop:ub}, we promised upper bounds on the complexity of elementary symmetric polynomials. 
The promise we now fulfil. 

\paragraph*{A quadratic upper bound in the homogeneous setting}

We want to show that for $d\in \{0,\dots, n\}$, $\esym{n}{d}$ can be simultaneously computed by a homogeneous circuit of size $O(n^2)$.
  
Note that % (we set $\esym{n,d}=0$ if $d>n$)
\[
  \esym{n}{d}(x_1, \ldots, x_n) = \esym{n-1}{d-1}(x_1, \ldots, x_{n-1}) \cdot x_n + \esym{n-1}{d}(x_1, \ldots, x_{n-1}).
\]
Hence, once we have computed $\esym{n-1}{d}$, $d\in \{0,\dots, n-1\}$, we can compute $\esym{n}{d}$, $d\in \{0,\dots, n\}$ using $O(n)$ extra gates. The overall complexity is quadratic.

\paragraph*{An almost linear upper bound in the non-homogeneous setting}

We want to show that $\esym{n}{d}$, $d\in \{0,\dots, n\}$, can be simultaneously computed by a non-commutative circuit of size $n\cdot \mathrm{poly}(\log n)$. 

The proof is the same as its commutative analog for elementary symmetric polynomials, see \cite{BS83} or the monograph by Burgisser et al. \cite[Chapters  2.1-2.3]{BCS97}. 
%In the non-scalar setting, however, the latter proof contains some highly non-trivial components. 

The main observation is that polynomial multiplication can be done efficiently. Let
\[ 
  f= \sum_{i=0}^n y_i t^i, \qquad g= \sum_{i=0}^n z_i t^i,  
\] 
where $f,g\in \F\inangle{y_0,\dots, y_n, z_0,\dots, z_n}[t]$. 
In other words, we assume that $t$ commutes with otherwise non-commuting variables $y_0,\dots, y_n$,$ z_0,\dots, z_n$. 
We view  $f,g$ as univariate polynomials in the variable $t$ with non-commutative coefficients. 
Then $fg= \sum_{i=0}^{2n} c_i t^i $ with $c_i=\sum_{j=0}^i y_j z_{i-j}$. 
Commutatively, the polynomials $c_0,\dots,c_{2n}$ can be simultaneously computed by a small circuit.
Indeed, if $\F$ contains sufficiently many roots of unity, one can obtain an $O(n\log n)$ circuit using Fast Fourier Transform; in other fields there are modification giving a circuit of size $O(n\log n\log\log n)$ {see \cite{SS71, BCS97}}. 
When counting only non-scalar product gates, this can be improved to $O(n)$ if $\F$ is sufficiently large.  
We observe that the same holds if the coefficients of $f,g$ do not commute. 
This is because the polynomials $c_k$ are bilinear in $y_0,\dots, y_n$,$ z_0,\dots, z_n$. 
Commutativity does not make a difference in this case (an exercise). 

Now consider the polynomial $h_n(t)=\prod_{i=1}^{n} (x_i + t) \in \F\inangle{x_1, \ldots, x_n}[t]$. 
Then one can see that $\esym{n}{d}(x_1,\dots,x_n)$ is the coefficient of $t^{n-d}$ in $h(t)$. 
The coefficients can be be recursively computed by first computing $\prod_{i=1}^{\ceil{n/2}} (x_i + t)$,  $\prod_{i=\ceil{n/2}+1}^n (x_i + t)$, and then combining the two by means of the fast polynomial multiplication above. 
This gives the claimed complexity.

\section{Open problems}
We end with two open problems. 

\begin{openproblem} 
  Find an explicit bivariate polynomial of degree $d$ which requires non-commutative homogeneous circuit of size superlinear in $d$
\end{openproblem}

\begin{openproblem} 
  Given a non-commutative monomial $\alpha$, can addition gates help to compute $\alpha$? 
\end{openproblem}

Observe that the bounds obtained in this paper are barely linear in $d$. Problem 1 simply asks for a quantitative improvement. A circuit with no addition gates is automatically homogeneous -- hence a negative answer to Problem 2 would allow to remove the homogeneity assumption in \autoref{thm:lin_lb_monomial}. 

\section*{Acknowledgement}
The first author thanks \href{https://www.cafedu.cz/en/}{Cafedu} for being such a nice place to work from.
The second author thanks Amir Yehudayoff for useful ideas on this topic which were exchanged in distant and joyous past.

\bibliographystyle{customurlbst/alphaurlpp}
\bibliography{references}

\end{document}